\documentclass{JHEP3} % 10pt is ignored!

\usepackage{epsfig,multicol}
\usepackage{amsmath, amssymb}

\usepackage{concmath, palatino}

\newcommand\fverb{\setbox\pippobox=\hbox\bgroup\verb}
\newcommand\fverbdo{\egroup\medskip\noindent%
            \fbox{\unhbox\pippobox}\ }
\newcommand\fverbit{\egroup\item[\fbox{\unhbox\pippobox}]}
\newbox\pippobox
\newcommand{\be}{\begin{equation}}
\newcommand{\ee}{\end{equation}}

\newcommand{\ba}{\begin{eqnarray}}
\newcommand{\ea}{\end{eqnarray}}

\newcommand{\la}{\longrightarrow}

\newcommand{\ads}{AdS_5\times S^5}

\title{The scaling function at strong coupling \\
from the quantum string Bethe equations}

\author{Matteo Beccaria\\
  Dipartimento di Fisica, Universita' del Salento,
  Via Arnesano, I-73100 Lecce\\
  INFN, Sezione di Lecce\\
  E-mail: \email{matteo.beccaria@le.infn.it}}

\author{Gian Fabrizio De Angelis\\
  Dipartimento di Fisica, Universita' del Salento,
  Via Arnesano, I-73100 Lecce\\
  INFN, Sezione di Lecce\\
  E-mail: \email{deangelis@le.infn.it}}

\author{Valentina Forini\\
  Dipartimento di Fisica, Universita' di Perugia,
  Via A. Pascoli, I-06123 Perugia\\
  INFN, Sezione di Perugia\\ and\\
  Humboldt-Universit\"{a}t zu Berlin, Institut f\"{u}r Physik, Newtonstra{\ss}e
  15, D-12489 Berlin\\
  E-mail: \email{forini@pg.infn.it,forini@physik.hu-berlin.de
}}

\preprint{HU-EP-07/06 }

\abstract{
We study at strong coupling the scaling function describing the large spin anomalous dimension of twist two operators
in ${\cal N}=4$ super Yang-Mills theory.
In the spirit of  AdS/CFT duality, it is possible to extract it from the string Bethe Ansatz equations
in the $\mathfrak{sl}(2)$ sector of the $\ads$ superstring. To this aim,
we present a detailed analysis of the Bethe equations by numerical and analytical methods. We recover several short string semiclassical
results as a check. In the more difficult case of the long string limit providing the scaling function, we analyze the strong coupling
version of the Eden-Staudacher equation, including the Arutyunov-Frolov-Staudacher phase. We prove that it admits a unique solution, at least in perturbation theory,
leading to the correct prediction consistent with semiclassical string calculations.
}


\begin{document}

\section{Introduction}
\label{Sec:Intro}

The $n$-gluon maximally helicity violating (MHV) amplitudes in planar ${\cal N}=4$ SYM obey very remarkable
iterative relations~\cite{Anastasiou:2003kj,Bern:2005iz,Bern:1997nh} suggesting solvability or even integrability of the maximally supersymmetric gauge theory.
The main ingredient of the construction is the so-called {\em scaling function} defined in
terms of the large spin anomalous dimension of leading twist operators in the gauge theory~\cite{Sterman:2002qn}.
The scaling function can be obtained by considering operators in the $\mathfrak{sl}(2)$ sector of the form
\be
\mbox{Tr}\ \left\{ {\cal D}^S Z^L + \mbox{permutations}\right\}.
\ee
The classical dimension is $S+L$, so $L$ is the twist, with minimal value $L=2$. The minimal anomalous dimension in this sector
is predicted to scale at large spin $S$ as
\be
\Delta -S = f(g)\,\log S + {\cal O}(S^0),
\ee
where the planar 't Hooft coupling is defined as usual by
\be
g^2 = \frac{\lambda}{16\,\pi^2},\qquad \lambda = N\,g_{YM}^2.
\ee
The one and two loops explicit perturbative calculation of $f(g)$ is described in~\cite{Gross:1973ju,Georgi:1951sr,Dolan:2000ut} and~\cite{Kotikov:2002ab,Kotikov:2003fb}.
Based on the QCD calculation \cite{Moch:2004pa}, the three-loop ${\cal O}(g^6)$ calculation is performed in~\cite{Kotikov:2004er,Kotikov:2005ne} by
exploiting the so-called trascendentality principle (KLOV).

In principle, one would like to evaluate the scaling function, possibly at all loop order by
Bethe Ansatz methods exploiting the conjectured integrability of SYM. This strategy has been
started in~\cite{Eden:2006rx}. In that paper, an integral equation providing $f(g)$ is
proposed by taking the large spin limit of the Bethe equations~\cite{Beisert:2005fw}. Its
weak coupling expansion disagrees with the four loop contribution. The reason of this
discrepancy is well understood. The Bethe Ansatz equations contain a scalar phase, the
dressing factor, which is not constrained by the superconformal symmetry of the model. Its
effects at weak-coupling show up precisely at the fourth loop order.

A major advance was done by Beisert, Eden and Staudacher (BES) in \cite{Beisert:2006ez}. In
the spirit of AdS/CFT duality, they considered the dressing factor at strong coupling. In
that regime, it has been conjectured a complete asymptotic series for the dressing
phase~\cite{Beisert:2006ib}. This has been achieved  by combining the tight constrains from
integrability, explicit 1-loop $\sigma$-model
calculations~\cite{Arutyunov:2004vx,Beisert:2005cw,Hernandez:2006tk,Freyhult:2006vr} and
crossing symmetry~\cite{Janik:2006dc}. By an impressive insight, BES proposed a
weak-coupling all-order continuation of the dressing. Including it in the ES integral
equation they obtained a new (BES) equation with a rather complicated kernel. The predicted
analytic four-loop result agrees with the KLOV principle. Very remarkably, an explicit and
independent perturbative 4-loop calculation of the scaling function appeared
in~\cite{Bern:2006ew}. In the final stage, the 4-loop contribution is evaluated numerically
with full agreement with the BES prediction.

This important result is one of the main checks of AdS/CFT duality. Indeed, a non trivial perturbative quantity is evaluated in the gauge theory
by using in an essential way input data taken from the string side.

As a further check, one would like to recover at strong coupling the asymptotic behavior of
the scaling function, as predicted by the usual semiclassical expansions of spinning string
solutions~\cite{Gubser:2002tv,Frolov:2002av,Frolov:2006qe}. Actually, the BES equation
passes this check, partly numerically \cite{Benna:2006nd} and partly by analytical
means~\cite{Alday:2007qf}. One could say that this is a check that nothing goes wrong if one
performs the analytic continuation of the dressing phase from strong to weak coupling.

From a different perspective, one would like to close this logical circle and check that the
same result is obtained in the framework of the quantum string Bethe equations proposed
originally in~\cite{Arutyunov:2004vx}. Indeed, it would be very nice to show that these
equations reproduce the scaling function in the suitable long string limit. Also, one
expects to find some simplifications due to the fact that only the first terms in the strong
coupling expression of the dressing must be dealt with. On the other side, the BES equation
certainly requires all the weak-coupling terms if it has to be extrapolated at large
coupling.

In this paper, we pursue this approach. As a first step, we study numerically the quantum Bethe Ansatz equations in the $\mathfrak{sl}(2)$
sector and check various results not directly related to the scaling function. Then, we work out the long string limit which is relevant to the
calculation of $f(g)$. From our encouraging numerical results, we move to an analytical study of a new version of the BES equation suitable
for the string coupling region. This equation has been first derived by Eden and Staudacher in~\cite{Eden:2006rx}
as a minor result. Indeed, it has been left over because the main interest was focusing on matching the weak-coupling 4-loop prediction.
However, we believe that it is a quite comfortable tool if the  purpose is that of reproducing the strong coupling behavior of the scaling function.
We indeed prove that the solution described in~\cite{Alday:2007qf} is the unique solution of the strong coupling ES equation.

The plan of the paper is the following. In Sec.~(\ref{Sec:nodressing}) we recall the Bethe
Ansatz equations valid in the $\mathfrak{sl}(2)$ sector of ${\cal N}=4$ SYM without and with
dressing corrections. In Sec.~(\ref{Sec:semiclassical}) we present various limits obtained
in the semiclassical treatment of the $\ads$ superstring. We present our results for short
and long string configurations. In Sec.~(\ref{Sec:SBES}) we analyze the strong coupling ES
equation building explicitly its solution and checking that it agrees with the result
of~\cite{Alday:2007qf}. We also investigate numerically the equation without making any
strong coupling expansion to show that the equation is well-defined.
Sec.~(\ref{Sec:conclusions}) is devoted to a summary of the presented results.

\section{Gauge Bethe Ansatz predictions for the scaling function without dressing}
\label{Sec:nodressing}

In the seminal paper~\cite{Eden:2006rx}, Eden and Staudacher (ES) proposed to study the
scaling function in the framework of the Bethe Ansatz for the $\mathfrak{sl}(2)$ sector of
${\cal N}=4$ SYM. The states to be considered in this rank-1 perturbatively closed sector
take the general form 
\be 
\mbox{Tr}\left\{ \left({\cal D}^{s_1}\,Z\right)\ \cdots
\left({\cal D}^{s_L}\,Z\right) \right\},\quad s_1 + \cdots + s_L = S. 
\ee 
They are
associated to the states of an integrable spin chain. The anomalous dimension $\Delta$ is
related to the chain energies by 
\be 
\label{eq:classquant}
\Delta = L+S+E_{L, S}(g)\, . 
\ee 
The all-loop
conjectured Bethe Ansatz equations valid for $E_{L, S}$ up to wrapping terms are fully
described in \cite{Staudacher:2004tk,Beisert:2005fw}. Some explicit tests are can be found
in \cite{Eden:2005bt,Zwiebel:2005er}. The Bethe Ansatz equations for the roots
$\{u_k\}_{1\le k \le S}$ are 
\be \label{eq:basl2} 
\left(\frac{x^+_k}{x^-_k}\right)^L =
\prod_{j\neq k}^S\frac{x^-_k-x^+_j}{x^+_k-x^-_j} \frac{\displaystyle 1-\frac{g^2}{x^+_k
x^-_j}}{\displaystyle 1-\frac{g^2}{x^-_k x^+_j}}, \qquad x^\pm_k =
x\left(u_k\pm\frac{i}{2}\right), 
\ee 
where we have defined the maps 
\be x(u) = \frac{u}{2}
\left(1+\sqrt{1-\frac{4\,g^2}{u^2}}\right),\qquad u(x) = x + \frac{g^2}{x}. 
\ee 
The
solutions of Eq.~(\ref{eq:basl2}) must obey the following  constraint to properly represent
single trace operators 
\be 
\prod_{k=1}^S \frac{x^+_k}{x^-_k} = 1. 
\ee 
The quantum part of
the anomalous dimension, {\em i.e.} the chain spectrum, is obtained from 
\be 
E_{L, S}(g) =
2\,g^2\sum_{s=1}^S\left(\frac{i}{x^+_k}-\frac{i}{x^-_k}\right). 
\ee 
Taking the large $S$
limit of the Bethe Ansatz equations, ES obtain the following representation of the scaling
function 
\be 
f(g) = 8\,g^2-64\,g^4\int_0^\infty dt\,\sigma(t)\,\frac{J_1(2\,g\,t)}{2\,g\,t},
\ee 
where $\sigma$ is the solution of the integral equation 
\be 
\sigma(t) =
\frac{t}{e^t-1}\left[\frac{J_1(2\,g\,t)}{2\,g\,t}-4\,g^2\int_0^\infty dt'\,K_m(2\,g\,t,
2\,g\,t')\,\sigma(t')\right], 
\ee 
with the (so-called {\em main}) kernel 
\be K_m(t, t') =
\frac{J_1(t)\,J_0(t')-J_0(t)\,J_1(t')}{t-t'}. 
\ee 
Notice that given $\sigma(t)$ we can
simply write $f(g) = 16\,g^2\,\sigma(0)$~\cite{Lipatov:2006aa}. These equations are
independent on the twist which drops in the large $S$ limit. This is important since the
scaling function is expected to be universal \cite{Belitsky:2006en,Eden:2006rx} and
therefore can be computed at large twist.

Unfortunately, the perturbative expansion of $f(g)$ disagrees at 4 loops with the explicit
calculation in the gauge theory. This is well known to be due to the missing contribution of
the dressing phase.

\subsection{Input from string theory: Dressing corrections}

The effect of dressing is discussed in \cite{Beisert:2006ez} to which we defer the reader
for general discussions about its origin and necessity. The Bethe equations are corrected by
a universal dressing phase according to 
\be \label{eq:basldress}
\left(\frac{x^+_k}{x^-_k}\right)^L = \prod_{j\neq k}^S\frac{x^-_k-x^+_j}{x^+_k-x^-_j}
\frac{\displaystyle 1-\frac{g^2}{ x^+_k x^-_j}}{\displaystyle 1-\frac{g^2}{ x^-_k x^+_j}}\
e^{2\,i\,\theta_{kj}}. 
\ee
The general perturbative expansion of the dressing phase
is~\cite{Arutyunov:2004vx,Beisert:2005wv}~\footnote{This general formula holds unchanged in
various deformations of the SYM theory~\cite{Frolov:2005ty,Beisert:2005he}, see for
ex.~\cite{Astolfi:2006is}.} 
\be 
\theta_{kj} = \sum_{r\ge 2}\sum_{\nu\ge 0}\sum_{\mu\ge \nu}
\left(g^2\right)^{r+\nu+\mu} \beta_{r, r+1+2\nu}^{(r+\nu+\mu)}\left[q_r(p_k)
q_{r+1+2\nu}(p_j) - (k\leftrightarrow j)\right], 
\ee 
where the higher order charges $q_r(p)$
are~\cite{Beisert:2004hm} 
\be q_r(p)
=\frac{2\sin\left(\frac{r-1}{2}p\right)}{r-1}\left(\frac{\sqrt{1+16\,g^2\sin^2\frac{p}{2}}-1}{4\,g^2\,\sin\frac{p}{2}}\right)^{r-1}.
\ee 
The first non trivial constant is $\beta^{(3)}_{2,3}\neq 0$. Indeed,
$\beta^{(2)}_{2,3}=0$ consistently with the 3-loops agreement with explicit perturbation
theory.

The proposed coefficients for the all-order weak-coupling expansion of the dressing
phase~\cite{Beisert:2006ib} are given in~\cite{Beisert:2006ez} (see also \cite{Gomez:2006mf}
and \cite{Kotikov:2006ts}). They read 
\be 
\beta_{r, r+1+2\nu}^{(r+\nu+\mu)} =
2(-1)^{r+\mu+1}\frac{(r-1)(r+2\nu)}{2\mu+1}\binom{2\mu+1}{\mu-r-\nu+1}\binom{2\mu+1}{\mu-\nu}\zeta(2\mu+1)
\ee 
and zero if $\mu-r-\nu+1 < 0$. This proposal is completely equivalent to a precise
modification of the kernel of the integral equation. It amounts to the replacement 
\be
K_m(t, t')\to K_m(t, t') + 2\, K_c(t, t'), 
\ee
where the dressing kernel can be written 
\be
K_c(t, t') = 4\,g^2\int_0^\infty dt''\,K_1(t,
2\,g\,t'')\,\frac{t''}{e^{t''}-1}\,K_0(2\,g\,t'', t'), 
\ee with 
\ba
K_0(t, t') &=& \frac{t\,J_1(t)\,J_0(t')  - t'\,J_1(t')\,J_0(t)}{t^2-(t')^2} , \\
K_1(t, t') &=& \frac{t'\,J_1(t)\,J_0(t') - t\,J_1(t')\,J_0(t)}{t^2-(t')^2} .
\ea
The modified integral equation can be exploited to compute the perturbative expansion of $f(g)$. Now, there is agreement with the 4 loop
explicit calculation. As we stressed in the Introduction, it is very remarkable that this weak coupling agreement is found with
various inputs from string theory.
In this sense, this is a powerful check of AdS/CFT duality.

\section{Strong coupling regime and the string Bethe equations}
\label{Sec:semiclassical}

As we explained, the BES equation is obtained by including in the ES equation an all-order weak-coupling expansion of the dressing phase.
This expansion comes from a clever combination of string theory inputs and constraints from integrability.
In our opinion, this is the essence of the integrability approach to AdS/CFT duality. As a consistency check, one would like to
recover from the BES equation the known semiclassical predictions valid in the $\ads$ superstring at large coupling.

There are indeed several limits that can be computed. The semiclassical limit is evaluated
in terms of the BMN-like~\cite{Berenstein:2002jq} scaled variables which are kept fixed as
$\lambda\to\infty$ 
\be 
\frac{E}{\sqrt{\lambda}},\ \frac{J}{\sqrt{\lambda}},\
\frac{S}{\sqrt{\lambda}}, 
\ee 
where $E, J, S$ are the semiclassical energy of a string
rotating in $S^5$ with angular momentum $J$ and spinning in $AdS_5$ with spin $S$. The
classical solution, and the first quantum corrections as well, are described in~\cite{Gubser:2002tv,Frolov:2002av,Frolov:2006qe}.

The simplest limits that can be considered are those describing short strings that do not probe $AdS$ regions with large curvature. We call them
\ba
\mbox{short-GKP},\qquad \frac{J}{\sqrt{\lambda}}\ll 1,  \frac{S}{\sqrt{\lambda}}\ll 1,\qquad  &:& E = \sqrt{J^2 + 2\,S\,\sqrt\lambda}, \\
\mbox{short-BMN},\qquad \frac{J}{\sqrt{\lambda}}\ \mbox{fixed},  \frac{S}{\sqrt{\lambda}}\ll 1,\qquad  &:& E = J + S\,\sqrt{1+\frac{\lambda}{J^2}}.
\ea

The scaling function is instead reproduced in the simplest long string limit which reads
\be
\mbox{long string},\qquad \frac{J}{\sqrt{\lambda}}\ll 1,  \frac{S}{\sqrt{\lambda}}\gg 1,\qquad  : E = S + f(\lambda)\,\log S .
\ee
In this limit, one can read the strong coupling behavior of the scaling function which is
\be
f_{\rm{string}}(g)=4\,g-\frac{3}{\pi}\,\log 2
+\cdots\ .
\ee

A first attempt to solve numerically the integral equation is described in
\cite{Benna:2006nd}. The BES equation is solved in a discrete series of Bessel modes and the
result is a numerical profile of the scaling function shown in Fig.~(1) of that paper. The
lower curve is the scaling function taking into account the conjectured dressing and well
reproduces  the string prediction.

Recently, the leading term of $f(g)$ has been obtained analytically in~\cite{Alday:2007qf}, including the analytic expression of the Bethe roots
densities in the strong coupling limit.

An intriguing feature of  Fig.~(1) in~\cite{Benna:2006nd} is that the strong coupling behavior starts very early, namely  at $g\simeq 1$. The long string behavior
should be visible at $S/\sqrt\lambda \gg 1$ which at $g=1$ means $S \gg 13$. One can ask if it is possible to explore numerically the Bethe equations
in the gauge theory up to $g\simeq 1$ and with $S = {\cal O}(100)$ to extract the scaling function. Actually, this is a hard task.
At $g\simeq 1$ the regime is not perturbative. The complete dressing should be resummed and it is not easy to do that, although some
very interesting results have been presented in \cite{Beisert:2006ez}.

An alternative {\em hybrid} approach would be that of taking the string Bethe equations with leading dressing. This should be enough to study
the leading terms of $f(g)$ at strong coupling. Of course, the problem is now that $g$ must be large and then $S$ must be unrealistically large
to deal with the numerical solution. However, not much is known
about the properties of the strong coupling dressing expansion. It is divergent, but possibly asymptotic. Therefore, it would be difficult to
estimate its accuracy at $g\simeq 1$.

In the next Section, we illustrate the detailed exploration of the above three limits.

\subsection{The String Bethe equations and their numerical solution}

For the subsequent analysis it is convenient to pass to the $p$ variables defining
\ba
p(u) &=& -i \log\frac{x^+(u)}{x^-(u)}, \\
x^\pm(p) &=& \frac{\displaystyle e^{\pm\frac{ip}{2}}}{4\sin\frac{p}{2}}\left(1+\sqrt{1+16\,g^2\sin^2\frac{p}{2}}\right), \\
u(p) &=& \frac{1}{2}\cot\frac{p}{2}\sqrt{1+16\,g^2\sin^2\frac{p}{2}}.
\ea
The  loop corrections to the energy are now
\be
E_{L, S}(g) = \sum_{k=1}^S \left(\sqrt{1+16\,g^2\sin^2\frac{p_k}{2}}-1\right),
\ee
where $\{p_k\}$ are obtained by solving the Bethe equations with dressing Eq.~(\ref{eq:basldress}). At strong coupling, we use the leading
dressing phase and write the string Bethe equations in logarithmic form as
\ba
i\,L\,p_k &=& \sum_{j\neq k}\left\{
\log\left(
\frac{x^-_k-x^+_j}{x^+_k-x^-_j}
\frac{\displaystyle 1-\frac{g^2}{x^-_k x^+_j}}{\displaystyle 1-\frac{g^2}{x^+_k x^-_j}}
\right) + \right. \\
&& \left. +  2\,i\,(u_k-u_j)\log\left(
\frac{\displaystyle 1-\frac{g^2}{x^-_k x^+_j}}{\displaystyle 1-\frac{g^2}{x^+_k x^+_j}}
\frac{\displaystyle 1-\frac{g^2}{x^+_k x^-_j}}{\displaystyle 1-\frac{g^2}{x^-_k x^-_j}}
\right)\right\}+2\,\pi\,i\,\varepsilon(p_k)
\ea
where $\varepsilon(x) = x/|x|$ and we have utilized the $p\to -p$ symmetry of the solution for the ground state as well as its known mode
numbers~\cite{Eden:2006rx}.

The numerical solution of the String Bethe equations is perfectly feasible. The techniques have already been illustrated in the
two compact rank-1 subsectors $\mathfrak{su}(2)$ and $\mathfrak{su}(1|1)$ as discussed in \cite{Beccaria:2006td,Beccaria:2007qx}.

First, we solve the equations at $\lambda=0$. This is the one-loop contribution. It is known exactly at $L=2$.
The Bethe roots are obtained,  for an even spin $S$,  as the roots
of the resolvent polynomial~\cite{Korchemsky:1995be,Eden:2006rx}
\be
\prod_s(u-u_s) \sim {}_3 F_2\left(-S, S+1, \frac{1}{2}-i\,u; 1, 1; 1\right).
\ee
The 1-loop energy can also be computed in closed form with the result
\be
E_{2, S} = 8\,g^2\,(\psi(S)-\psi(1))\sim 8\,g^2\,\log\,S + \cdots\la f(g) = 8\,g^2 + \cdots.
\ee
The roots are real and symmetrically distributed around zero. This features are also true for the ground state at
arbitrary finite twist.

At $L>2$, 
we use the solution at $L=2$ as the starting point for the numerical root finder. Then, we increase $g$. At each step, we use a linear
extrapolation of the previous solutions to improve the guess of the new solution. This procedure is quite stable and allows
to explore a wide range of $L, S, g$ values. Notice that changing the twist $L$ is trivial since the complexity of the equations
does not change.

\subsection{Short string in the GKP limit}

As a first numerical experiment we fix $L$ and $S$ and increase $g$ up to large values where the equations are reliable. This is the short string limit
where the $AdS$ geometry is approximately flat. One expects to recover the Gubser-Klebanov-Polyakov law $\Delta\sim g^{1/2}$~\cite{Gubser:1998bc}.
Solving the equations, we indeed verify that the Bethe momenta have the asymptotic scaling $p_k\sim g^{-1/2}$. This is clearly illustrated in Fig.~(\ref{fig:scaling}).

Despite the non trivial distribution of the Bethe roots, it is straightforward to compute the anomalous dimension at large $g$.
To do that, we first write the Bethe equations in the form
\be
i\,L\,p_k = 2\,\pi\,i\,\varepsilon(p_k) + \sum_{j\neq k} S_{kj}.
\ee
At large $g$ we can write
\be
p_k = \frac{\alpha_k}{g^{1/2}} + \frac{\alpha_k'}{g} + \cdots.
\ee
The Bethe momenta can be divided into a set $P$ with $\alpha_k>0$ and a symmetric set obtained by flipping $p\to -p$.
The expansion of the energy is
\ba
E &=& \sum_k\left(2\,g^{1/2}\,|\alpha_k| + 2\,\varepsilon(\alpha_k)\,\alpha'_k-1\right) + \cdots = \\
&=& 4\,g^{1/2}\,\sum_{k\in P}\alpha_k + 4\,\sum_{k\in P}\alpha'_k - S. \ea
From the Bethe
equations we have 
\be i\,L\,\sum_{k\in P} p_k = i\,\pi\,S + \sum_{k\in P}\sum_{j\not\in P}
S_{kj}. 
\ee
 where we exploited $S_{ij} = -S_{ji}$. Each Bethe momentum $p_j$ with $j\not\in
P$ can be written as $-p_{j'}$ with $j'\in P$. Hence we can write 
\be i\,L\,\sum_{k\in P}
p_k = i\,\pi\,S + \sum_{j,k\in P} S(p_k, -p_j). 
\ee
At large $g$ 
\be S(p_k, -p_j) =
-2\,i\,\left(\alpha_k\,\alpha_j + \frac{1}{g^{1/2}}(\alpha'_k\,\alpha_j +
\alpha_k\,\alpha'_j)\right) + \cdots 
\ee
The leading order Bethe equations give 
\be
i\,\pi\,S-2\,i\sum_{k,j\in P} \alpha_k\,\alpha_j = 0\ \la\ \sum_{k\in P}\alpha_k =
\left(\frac{\pi\,S}{2}\right)^{1/2}. 
\ee
The next-to-leading terms are 
\be i\,L\,\sum_{k\in
P}\alpha_k = -2\,i\,\sum_{j,k\in P} (\alpha'_k\,\alpha_j + \alpha_k\,\alpha'_j) =
-4\,i\,\sum_{k\in P} \alpha_k\,\sum_{j\in P} \alpha'_j, 
\ee
\be \qquad  \la\ \sum_{k\in P}
\alpha'_k = -\frac{L}{4}. 
\ee
In conclusion, by plugging these results in the expression for
$\Delta$, Eq.~(\ref{eq:classquant}), we find 
\be \label{eq:gkp} 
\Delta = \sqrt{2\,S}\,\lambda^{1/4}
%-L-S 
+ {\cal O}(\lambda^{-1/4}). 
\ee
The subleading term cancels the classical contribution leaving a
pure $\lambda^{1/4}$ behavior. This can be compared with the flat limit in the semiclassical
approximation that reads for $J/\sqrt\lambda$, $S/\sqrt\lambda \ll 1$ 
\be E_{\rm string} =
\sqrt{J^2 + 2\,S\,\sqrt{\lambda}} 
\ee
 with full agreement.

\subsection{Short string in the BMN limit}

If we keep $S$ fixed and increase $\lambda$ with $\lambda/L^2$ fixed, we can reach the BMN limit~\cite{Berenstein:2002jq}. This is numerically very easy because
$L$ enters trivially the equations. Fig.~(\ref{fig:bmn1}) shows the convergence to the BMN limit when $L$ is increased from 10 to 100 and $S$ is fixed
at $S=4$. The various curves clearly approach a limiting one. This is very nice since it is an explicit show of how the BMN regimes sets up.
Fig.~(\ref{fig:bmn2}) shows the limiting curves for $S=4, 6, 8$ at very large $L=10^4$. The three curves are perfectly fit by the expected law
\be
E = L + S\,\sqrt{1+\frac{\lambda}{L^2}}.
\ee

\subsection{Long string limit and the scaling function}

The previous pair of tests in the (easy) short string limits is a clear illustration that the numerical solution of the Bethe equations is reliable.

The slow string limit is much more difficult. We begin with a plot of the energy at fixed
twist $L=6$ and increasing spin from 10 to 60. It is shown in Fig.~(\ref{fig:slow1}). Each
curve bends downward as $g$ increases, since it ultimately must obey the $g^{1/2}$ law.
However, at fixed $g$, when $S$ increases the energy increases slowly eventually following
the $\log S$ law. We attempted an extrapolation at $S\to\infty$ at each $g$. In
Fig.~(\ref{fig:der}) we show our estimate for the derivative of the scaling function, by
fitting the data $(S_1-S_2)$ with $S_1 < S< S_2$. We also show the analytical prediction
$4$. It seems to be roughly reproduced as soon as $g\gtrsim 1$.

The above "dirty" numerical procedure shows that it is reasonable to expect that the quantum
string Bethe equations are able to capture the correct strong coupling behavior of the
scaling function. However, the above extrapolation has a high degree of arbitrariness,
especially concerning the fitting function employed to estimate the $S\to\infty$ limit.
Also, one would like to go to quite larger $g$ requiring a huge number of Bethe roots, equal
indeed to the spin $S$. In practice, as it stands, the numerical investigation could hardly
be significantly improved.

For all these reasons, in the final part of this paper we explore an equation analogous to
the BES equation, but derived for the string Bethe equation, at least with the leading order
dressing phase.

\section{The strong coupling ES equation}
\label{Sec:SBES}

The inclusion of the AFS phase~\cite{Arutyunov:2004vx}  in the ES equation
\ba
\sigma(t) &=& \frac{t}{e^t-1}\left[\frac{J_1(2\,g\,t)}{2\,g\,t}-4\,g^2\,\int_0^\infty dt'\, K(2\,g\,t, 2\,g\,t')\  \sigma(t')\right], \\ \nonumber \\
K(t,t') &\equiv& K_m(t, t') = \frac{J_1(t) \,J_0(t')-J_1(t')\, J_0(t)}{t-t'},
\ea
is straightforward and has been described in full details in \cite{Eden:2006rx}. The resulting equation reads
\ba
\sigma(t) &=& \frac{t}{e^t-1}\left[\frac{J_1(2\,g\,t)}{2\,g\,t}+2\,g\,\frac{J_2(2\,g\,t)}{2\,g\,t}-4\,g^2\,\int_0^\infty dt'\, \widetilde{K}(2\,g\,t, 2\,g\,t')\ \sigma(t')\right], \\
\nonumber \\
\widetilde{K}(t,t') &=& K(t,t') + 2\,g\, K_{\rm AFS}(t, t'), 
\ea 
where the AFS kernel is 
\be
K_{\rm AFS}(t, t') = \frac{t\,(J_2(t)\, J_0(t')-J_2(t')\, J_0(t))}{t^2-(t')^2}. 
\ee 
It is
important to remark that both the main and AFS kernels admit a simple expansion as series of
products of two Bessel functions. For the main kernel, it is well known that 
\be K(t,t') =
\frac{2}{t\,t'}\sum_{n\ge 1} n\,J_n(t)\,J_n(t'). 
\ee
For the AFS kernel we have
\footnote{This expansion can be proved by starting from the identity 
\be J_2(t)\,
J_0(t')-J_2(t')\, J_0(t) = \frac{2}{t\,t'}\left[t'\, J_1(t)\, J_0(t')-t\, J_0(t)\,
J_1(t')\right]. 
\ee
The result Eq.~(\ref{eq:afsexp}) follows immediately from the expansions
reported in the appendices of~\cite{Beisert:2006ez}. } 
\be \label{eq:afsexp} \frac{J_2(t)\,
J_0(t')-J_2(t')\, J_0(t)}{t^2-(t')^2} = \frac{8}{t^2\,(t')^2}\sum_{n\ge 1} n\,
J_{2n}(t)\,J_{2n}(t'). 
\ee
We change variables to put the equation in a somewhat simpler
form and define 
\be s(t) = \frac{e^t-1}{t}\,\sigma(t),\qquad h(\tau) =
s\left(\frac{\tau}{2g}\right). 
\ee
The strong coupling ES equation for $h$ is then 
\be
h(\tau) = \frac{J_1(\tau)}{\tau}+2\,g\,\frac{J_2(\tau)}{\tau}-\int_0^\infty d\tau'\,
\widetilde{K}(\tau, \tau')\, \frac{\tau'}{e^{\tau'/(2\,g)}-1}\ h(\tau'). 
\ee
This equation
is expected to be reliable at strong coupling. It should reproduce the leading term in the
scaling function $f(g) = 8\,g^2\, h(0)$. With the conventions adopted in this part of the
paper, this means $h(0) = 1/(2g) +\mbox{subleading}$.

\subsection{Leading order at strong coupling}

We are interested in the large $g$ limit with $\tau$ fixed. The leading terms are obtained by expanding
%\be
%\frac{J_2(\tau)}{\tau}-\int_0^\infty d\tau'\, K_{\rm AFS}(\tau, \tau')\, \frac{\tau'}{e^{\tau'/(2\,g)}-1}\ h(\tau') = 0.
%\ee
%At leading order
\ba
h(\tau) &=& \frac{1}{g}\,v(\tau) + {\cal O}\left(\frac{1}{g^2}\right), \\
\frac{\tau}{e^{\tau/(2\,g)}-1} &=& 2\,g+ {\cal O}\left(1\right). \ea Taking the terms with
the leading power of $g$ we find that $v(\tau)$ satisfies the remarkably simple equation 
\be
\label{eq:strong} \int_0^\infty d\tau'\, K_{\rm AFS}(\tau, \tau') \ v(\tau') =
\frac{J_2(\tau)}{2\tau}. 
\ee
Taking into account the expansion of the AFS kernel, the
equation can be written in the equivalent form 
\be \int_0^\infty d\tau\,
\frac{J_{2n}(\tau)\, v(\tau)}{\tau^2} = \frac{1}{16}\,\delta_{n,1}. 
\ee
Following the
approach of \cite{Benna:2006nd,Alday:2007qf} we expand the solution in a Neumann series of
Bessel functions 
\be \label{eq:class} 
v(\tau) = \sum_{k\ge 1} c_k\,\frac{J_k(\tau)}{\tau},
\ee
and now the equation reads 
\be \label{eq:stronglinear} \sum_{k\ge 1} H_{n\,k}\,c_k =
\frac{1}{16}\,\delta_{n,1},\qquad H_{n\, k} = \int_0^\infty\frac{J_{2n}(x)\,J_k(x)}{x^3} dx.
\ee
Now, the following question arises: Is this equation a constraint or does it determine a
unique solution $v(\tau)$? As a first step, we prove that the solution of
\cite{Alday:2007qf} is indeed a solution to the above equation. This solution reads 
\be
\label{eq:alday} c_{2n} = (-1)^{n+1}\frac{\Gamma(n+1/2)}{\Gamma(n)\Gamma(1/2)},\qquad
c_{2n+1} = (-1)^{n}\frac{\Gamma(n+3/2)}{\Gamma(n+1)\Gamma(1/2)}. 
\ee
Now, the detailed
values of the matrix elements $H_{n\,k}$ are 
\be k\ \rm {even}:\ \ \ H_{n, k} =
\left\{\begin{array}{cc}
\displaystyle\frac{1}{32\,n\,(n+1)\,(2n+1)}, &\quad k=2n+2,\\
\displaystyle\frac{1}{8\,n\,(2n+1)\,(2n-1)}, &\quad k=2n,\\
\displaystyle\frac{1}{32\,(2n-1) n (n-1)}, &\quad k=2n-2,\\
0, & \quad \mbox{otherwise}.
\end{array}\right.
\ee
and
\be
k = 1+2p,\ :\ \ \ H_{n,1+2p} = \frac{16}{\pi}\frac{(-1)^{n+p}}{
(4n^2-(2p+3)^2)(4n^2-(2p+1)^2)(4n^2-(2p-1)^2)}
\ee
The linear equations Eq.~(\ref{eq:stronglinear}) are then (${\cal E}$ stands for {\em even Bessel})
\be
{\cal E}_n = 0,\qquad n\ge 1,
\ee
where
\be
\label{eq:linear}
{\cal E}_n = (1-\delta_{n,1})\,\frac{c_{2n-2}}{32(2n-1) n (n-1)}+
\frac{c_{2n}}{8n(2n+1)(2n-1)} +
\frac{c_{2n+2}}{32n(n+1)(2n+1)} +
\ee
$$
 + \sum_{p\ge 0} c_{2p+1} \, H_{n, 1+2p} - \frac{1}{16}\,\delta_{n, 1}.
$$
These equations are indeed satisfied by the solution defined through Eq.~(\ref{eq:alday}). This can be checked by
evaluating the  infinite sum in closed form by the Sommerfeld-Watson transformation methods.
For instance the five terms in Eq.~(\ref{eq:linear}) read at $n=1,2,3$
\ba
{\cal E}_1 &=& 0 + \frac{1}{48}-\frac{1}{256}+\frac{35}{768} -\frac{1}{16} = 0, \\
{\cal E}_2 &=& \frac{1}{384} - \frac{1}{320}+\frac{1}{1024}-\frac{7}{15360} - 0 = 0,\\
{\cal E}_3 &=& -\frac{1}{1280} + \frac{1}{896}-\frac{5}{12288}+\frac{311}{430080} -0 = 0.
\ea
However, this is not the unique solution of Eq.~(\ref{eq:strong}). As is quite usual,
the straightforward strong coupling limit of Bethe equations does not determine completely the solution which is
fixed by the tower of subleading corrections. A similar difficulty is explained in details in~\cite{Alday:2007qf}.

For instance, a second solution of Eq.~(\ref{eq:stronglinear}) is
\be
c_{2\,k} = 2\,k\,(-1)^{k+1},\qquad c_{2\,k+1} = 0.
\ee
Indeed, this produces the remarkably simple solution
\be
v(\tau) = \sum_{k\ge 1}2\,k\,(-1)^{k+1} \frac{J_{2\,k}(\tau)}{\tau} \equiv \frac{1}{2}\, J_1(\tau).
\ee
Notice that, before summing the series, this second solution is precisely of the general class Eq.~(\ref{eq:class}).
As a check, we have indeed
\be
\int_0^\infty d\tau\, \frac{J_{2n}(\tau)\, J_1(\tau)}{2\,\tau^2} = -\frac{1}{8\,\pi}\frac{\sin\,n\,\pi}{n^2\,(n^2-1)} =
\frac{1}{16}\ \delta_{n,1}\quad \mbox{for}\ n=1, 2, \dots
\ee
In practice, we still need the equal-weight condition on the even/odd Bessel functions contained in the solution~\cite{Alday:2007qf}.

To find a unique solution, we must examine the next orders in the strong coupling expansion.
Indeed, the next orders provide both equations for the various subleading corrections to the solution $v(\tau)$ {\em and}
constraints on the previous contributions. This is due to the fact that the AFS kernel $K_{\rm AFS}(\tau, \tau')$
is a function of $\tau$ expressed as a Neumann series of purely {\em even} Bessel functions. The {\em odd} Bessel functions
provide the above mentioned constraints as we now illustrate.

\subsection{NLO order at strong coupling}

Let us work out the constraints from the subleading correction. If we take into account the next terms in the expansion and write
\ba
h(\tau) &=& \frac{1}{g}\,v^{(0)}(\tau) +\frac{1}{g^2}\,v^{(1)}(\tau) + \cdots, \\
\frac{\tau}{e^{\tau/(2\,g)}-1} &=& 2\,g-\frac{\tau}{2} + \cdots
\ea
we find the following equation for $v^{(1)}$
\ba
\lefteqn{
\int_0^\infty d\tau'\ K_{\rm AFS}(\tau, \tau') \ v^{(1)}(\tau') = }&& \\
&& = \int_0^\infty d\tau'\,\left(\frac{1}{4} K_{\rm AFS}(\tau, \tau')\,\tau'-\frac{1}{2}\,K(
\tau, \tau')\right)\,v^{(0)}(\tau') + \frac{J_1(\tau)}{4\,\tau}.
\ea
To simplify, we exploit
\ba
\lefteqn{\frac{1}{4} K_{\rm AFS}(\tau, \tau')\,\tau'-\frac{1}{2}\,K(
\tau, \tau') = } && \\
&& = \frac{2}{\tau\,\tau'}\sum_{n\ge 1} n\,J_{2n}(\tau)\,J_{2n}(\tau')- \frac{1}{\tau\,\tau'}\sum_{n\ge 1} n\,J_{n}(\tau)\,J_{n}(\tau') = \\
&& = - \frac{1}{\tau\,\tau'}\sum_{n\ge 0} (2\,n+1)\,J_{2\,n+1}(\tau)\,J_{2\,n+1}(\tau').
\ea
Hence, the equation can be written
\be
\int_0^\infty d\tau'\ K_{\rm AFS}(\tau, \tau') \ v^{(1)}(\tau') =  \frac{J_1(\tau)}{4\,\tau} + \sum_{n\ge 0} (2\,n+1)\,\beta_n\,\frac{J_{2\,n+1}(\tau)}{\tau}
\ee
where, explicitly,  the constants $\beta_n$ are given by
\ba
\beta_n &=& -\sum_{k\ge 1} c_k\,G_{n\,k}, \\
G_{n\,k} &=& \int_0^\infty \frac{J_{2\,n+1}(x)\,J_{k}(x)}{x^2}\,dx.
\ea
The relevant integrals are
\be
\int_0^\infty \frac{J_{2\,n+1}(x)\,J_{2\,k}(x)}{x^2}\,dx =
\left\{\begin{array}{cc}
\displaystyle\frac{1}{8\,n\,(2\,n+1)}, &\quad k=n,\\
\displaystyle\frac{1}{8\,(n+1)\,(2\,n+1)}, &\quad k=n+1,\\
0, & \quad \mbox{otherwise}.
\end{array}\right.
\ee
and
\ba
F_{n\,k} &=& \int_0^\infty \frac{J_{2\,n+1}(x)\,J_{2\,k-1}(x)}{x^2}\,dx = \nonumber \\
&=& \frac{4\,(-1)^{n+k}}{\pi}\frac{1}{(4\,k^2-(2\,n+1)^2)(2\,k+2\,n-1)(2\,k-2\,n-3)}
\ea
Equating to zero the coefficients of the odd Bessel functions $J_{2n+1}(\tau)$ we obtain the constraint on $v^{(0)}$
\be
{\cal O}_n = 0,\qquad n\ge 0,
\ee
where
\be
{\cal O}_n = (1-\delta_{n,0}) \frac{c_{2n}}{8n(2n+1)} + \frac{c_{2n+2}}{8(n+1)(2n+1)} + \sum_{k\ge 1} F_{n\,k}\,c_{2\,k-1}-\frac{1}{4}\,\delta_{n,0}.
\ee
These conditions are linear combinations of the previous equations. Indeed, one can check that
\be
{\cal E}_n = \frac{1}{4\,n}({\cal O}_n + {\cal O}_{n-1}).
\ee
So this constraint adds nothing new and, in particular, is satisfied by the Alday's solution~\cite{Alday:2007qf}. Looking at the even Bessel functions,
we obtain the homogeneous equation
\be
\int_0^\infty\frac{J_{2n}(\tau)\,v^{(1)}(\tau)}{\tau^2}\,d\tau = 0.
\ee
which admits the consistent solution $v^{(1)} = 0$.

\subsection{NNLO order at strong coupling}

The next order in the $1/g$ expansion is what we need to fix uniquely $v^{(0)}$. Expanding as before and writing now
\ba
h(\tau) &=& \frac{1}{g}\,v^{(0)}(\tau) +\frac{1}{g^2}\,v^{(1)}(\tau) + \frac{1}{g^3}\,v^{(2)} + \cdots, \\
\frac{\tau}{e^{\tau/(2\,g)}-1} &=& 2\,g-\frac{\tau}{2} + \frac{\tau^2}{24\,g} + \cdots
\ea
we find the following equation for $v^{(2)}$
\be
4\,\int_0^\infty K_{\rm AFS}(\tau, \tau')\,v^{(2)}(\tau') \,d\tau' =
\ee
$$
= \int_0^\infty\left\{\frac{1}{2}\tau'\,K(\tau, \tau')-\frac{1}{12}(\tau')^2
K_{\rm AFS}(\tau, \tau')\right\}\,v^{(0)}(\tau') \,d\tau'-v^{(0)}(\tau).
$$
The odd Bessel functions give a constraint independent on $v^{(2)}$. To compute it, we need
the integrals (in our relevant range of values for $n, k$) 
\be
\int_0^\infty\frac{J_{2k+1}(x)\,J_{2n}(x)}{x}\,dx =
\frac{2}{\pi}\frac{(-1)^{n-k}}{(1+2k)^2-4n^2} 
\ee
and 
\be
\int_0^\infty\frac{J_{2k+1}(x)\,J_{2n+1}(x)}{x}\,dx = \frac{1}{2(1+2k)}\,\delta_{n,k}. \ee
One obtains immediately the crucial relation \be \frac{c_{2k+1}}{2k+1} =
\frac{4}{\pi}\sum_{n\ge 1}c_{2n}\,\frac{(-1)^{n-k}}{(1+2k)^2-4n^2}. 
\ee
This relation
permits to write all odd coefficients $c_{2k+1}$ in terms of the even ones. Substituting
this relation in the truncated versions of the basic conditions 
\be 
{\cal E}_n = 0, 
\ee
one
obtains a well-posed problem converging rapidly to the solution ~\cite{Alday:2007qf} without
any \emph{a priori} condition on the solution. For instance, by truncating the problem
dropping all $c_{2n}$ with $n>N$, we find the table Tab.~(\ref{tab:1}) of values for $c_2$.
A simple polynomial extrapolation to $N\to\infty$ provides the correct limit $c_1 = c_2 =
0.5000(1)$.

Hence, the strong coupling expansion is well defined and the leading solution is unique. Of course, it is
the one described in  ~\cite{Alday:2007qf}.

\TABLE{
\begin{tabular}{|| c|cccccccc || }
\hline
$N$   & 10 & 20 & 30 & 40 & 50 & 60 & 70 & 80 \\
$c_2$ & 0.5665 & 0.5386 & 0.5273 & 0.5211 & 0.5173 & 0.5146 & 0.5127 & 0.5112 \\
\hline
\end{tabular}
\caption{Coefficient $c_2$ from the truncated full rank linear problem.}
\label{tab:1}
}

\subsection{Numerical integration of the strong coupling ES equation}

To summarize, we have shown that the strong coupling ES equation is consistent with the results of \cite{Alday:2007qf}. In that paper,
it was crucial to fix the relative weights of the even/odd Bessel functions appearing in the general solution. These weights
were shown to be more than an Ansatz. They are encoded in the full equation before expanding at strong coupling. Alternatively, they can be
derived by analyzing the next-to-leading and $\mbox{next}^2$-to-leading corrections.

As a final calculation and check, we provide the results from a numerical investigation without
any strong coupling expansion to see how the correct strong coupling solution arises. This can be done along the lines
illustrated in~\cite{Benna:2006nd,Alday:2007qf}.
We start again from the Neumann expansion
\be
h(\tau) = \sum_{k\ge 1} c_k\,\frac{J_k(\tau)}{\tau},
\ee
and arrive at the infinite dimensional linear problem
\be
c_n = \delta_{n, 1} + 2\,g\,\delta_{n, 2}-2\,n\sum_{k\ge 1} Z_{n\,k}^{(1)}\,c_k-8\,g\,\left\{\begin{array}{cc}
0, &\quad n\,\mbox{odd} \\
n\,\sum_{k\ge 1} Z_{n\,k}^{(2)}\,c_k, &\quad  n\,\mbox{even}
\end{array}\right.
\ee
where
\be
Z_{n\,k}^{(p)}(g) = \int_0^\infty \frac{J_n(x)\,J_k(x)}{x^p\,(e^\frac{x}{2\,g}-1)}.
\ee
As in \cite{Benna:2006nd,Alday:2007qf} we can truncate this equation setting $c_k=0$ for $k> N$. The solution should be reliable for $g\lesssim N$.
We considered $2<g<30$ and $N=50$. The best fit to the leading term in $f(g)$ gives
\ba
f(g) &=& c\,g + \mbox{subleading}, \\
c &=& 4.000(1).
\ea
This confirms that the strong coupling ES equation has a unique solution with the correct large $g$ limit.
Of course, as $g\to\infty$ it is nothing but Eq.~(\ref{eq:alday}).

\section{Conclusions}
\label{Sec:conclusions}

In this paper, we have considered several properties of the quantum string Bethe equations in the $\mathfrak{sl}(2)$
sector with the leading strong coupling dressing, {\em i.e.} the AFS phase.
We have performed a numerical investigation of the equations showing that their analysis is quite feasible. As an interesting result,
we have repeated the calculation of the GKP limit of the anomalous dimensions as for the highest excited states in the compact
rank-1 subsectors $\mathfrak{su}(2)$ and $\mathfrak{su}(1|1)$.
Also, we have been able to observe the setting of the BMN scaling regime by reproducing the plane wave energy formula at fixed spin and large
twist.
In the case of the long string regime, we have been able to provide numerical evidences for a scaling function exhibiting an early strong
coupling behavior as expected from the numerical solution of the BES equation.

Motivated by these results, we have analyzed analytically and perturbatively at strong
coupling an almost trivially modified version of the BES equation with the very simple
strong coupling dressing~\cite{Arutyunov:2004vx}. In particular, we have proved that this
equation admits, as it should, a unique solution for the asymptotic Bethe root (Fourier
transformed) density in full agreement with existing results.

While this work was under completion, the paper~\cite{Kostov:2007kx} presented an analysis
partially overlapping with our results. That paper derives an integral equation for the
Bethe root density taking into account the dressing at strong coupling and is based on a
novel integral representation of the dressing kernel. We hope that the two alternative
approaches will turn out to be useful in computing the one-loop string correction to the
large $g$ scaling function. Indeed, this interesting contribution has been checked
numerically in \cite{Benna:2006nd} but it still evades an analytical confirmation.

Hopefully, these various efforts might give insight on the general structure
of the dressing phase as well as on the role of the asymptotic Bethe equations in an
exact description of the planar spectrum~\cite{Schafer-Nameki:2006ey}. Significative studies
about finite size effects~\cite{Ambjorn:2005wa} and corrections that arise in a finite
volume to the magnon dispersion relation at strong
coupling~\cite{Arutyunov:2006gs}, see also \cite{Astolfi:2007uz}, as well as the recent
observation~\cite{Rej:2007vm} that the dressing phase could originate from the elimination
of "novel" Bethe roots, strongly demand a deeper understanding.

\acknowledgments We thank D. Serban and M. Staudacher for very useful discussions and
comments. M. B. also thanks G. Marchesini for conversations about the properties of twist-2
anomalous dimensions at finite spin $S$. The work of V. F. is supported in part by the PRIN
project 2005-24045 "Symmetries of the Universe and of the Fundamental Interactions" and by
DFG Sonderforschungsbereich 647 "Raum-Zeit-Materie".

%\FIGURE{
%\epsfig{file=./Figures/benna.eps, width=14cm}
%\bigskip\bigskip
%\label{fig:benna}
%\caption{Scaling function from \cite{Benna:2006nd}. The lower curve is the relevant one, taking into account the all-order weak
%coupling expansion of the dressing phase.}
%}

\newpage
\FIGURE{
\epsfig{file=./scaling.eps, width=14cm}
\bigskip\bigskip
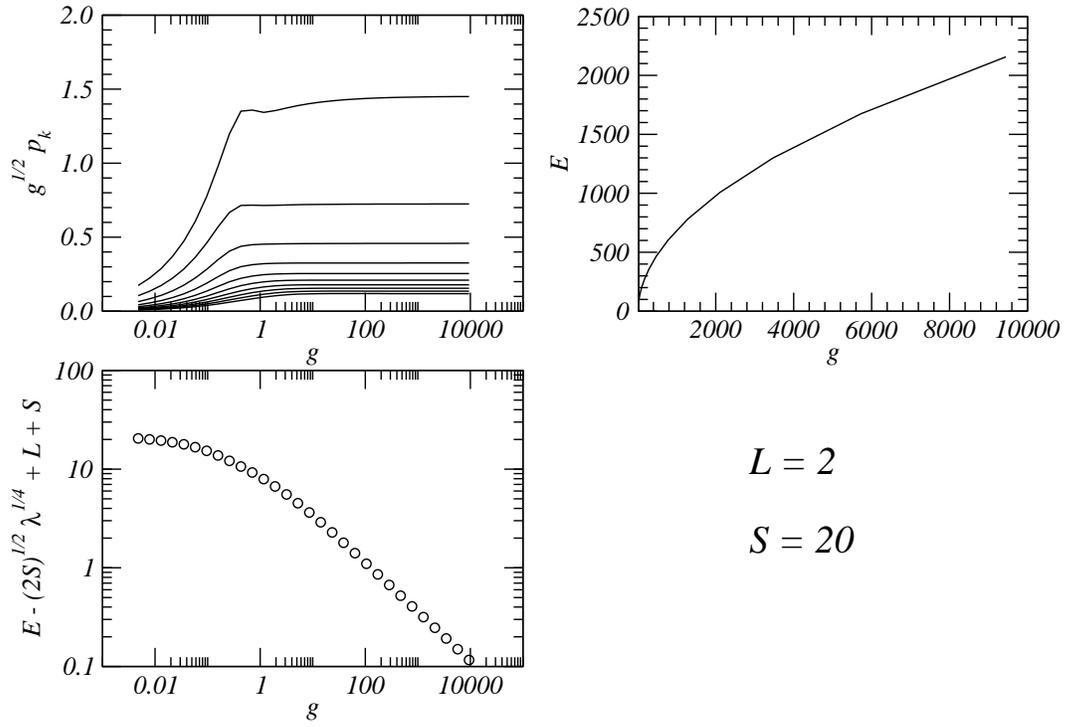
\caption{Numerical solution in the short-GKP limit, {\em i.e.} at $L, S$ fixed and
increasing $g$. We show the $g^{-1/2}$ behaviour of the Bethe momenta, the computed energy
with its $g^{1/2}$ profile, and the (vanishing) difference with analytic strong coupling
expansion Eq.~(\ref{eq:gkp}).} \label{fig:scaling} }

\newpage
\FIGURE{
\epsfig{file=./BMN1.eps, width=14cm}
\bigskip\bigskip
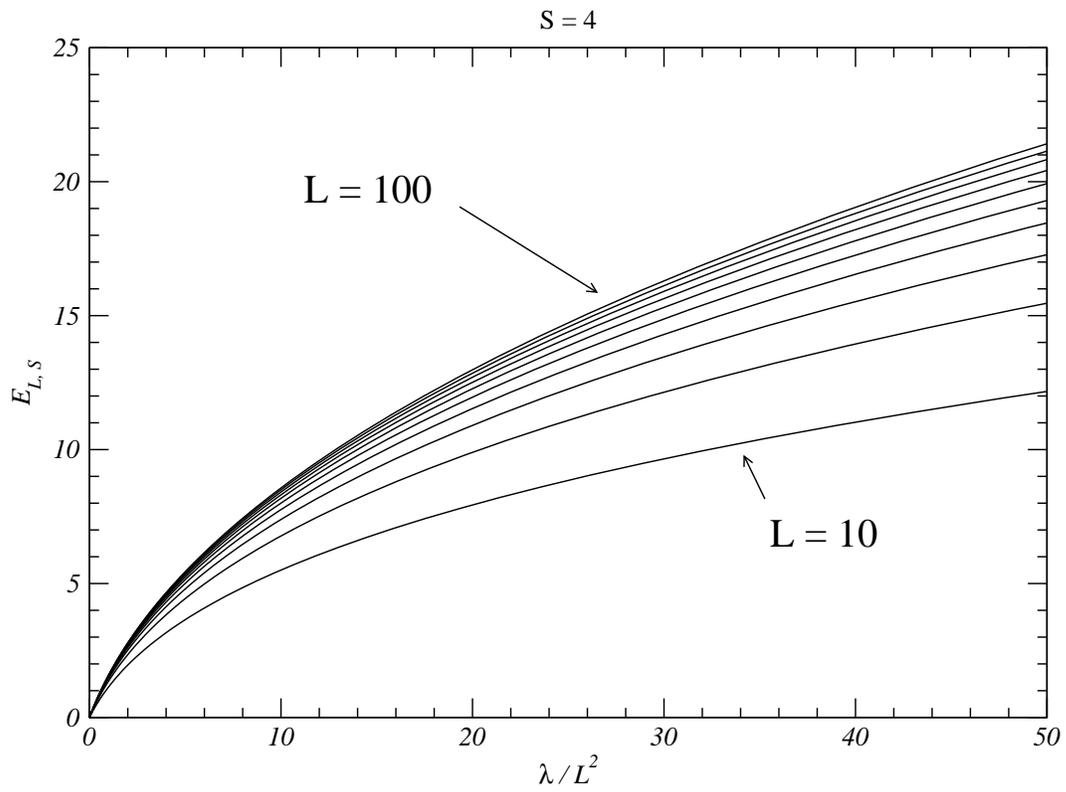
\caption{Short-BMN limit. Convergence to the BMN limit at $L, \lambda\to\infty$ with $\lambda/L^2$ fixed.}
\label{fig:bmn1}
}

\newpage
\FIGURE{
\epsfig{file=./BMN2.eps, width=14cm}
\bigskip\bigskip
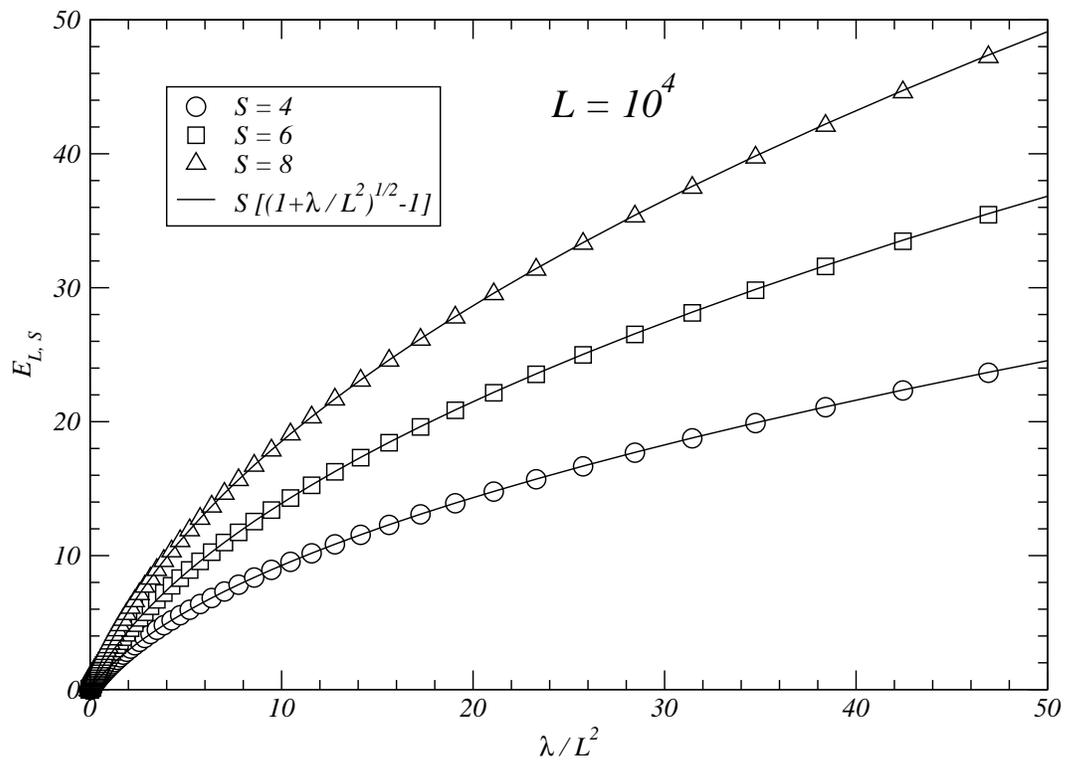
\caption{Short-BMN limit. Check of the semiclassical string prediction $E_{\rm string} = L + S\sqrt{1+\lambda/L^2}$.
We plot the quantum part $E_{L, S}$ computed by the Bethe Ansatz.}.
\label{fig:bmn2}
}

\newpage
\FIGURE{
\epsfig{file=./Slow1.eps, width=14cm}
\bigskip\bigskip
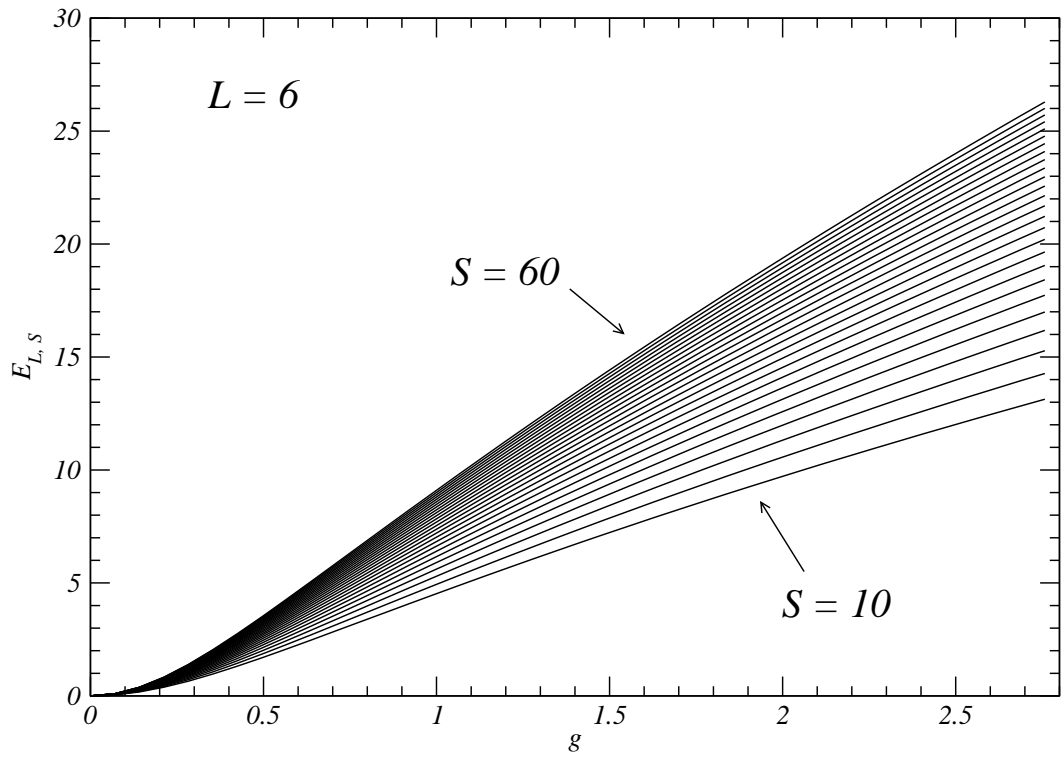
\caption{Long string limit. Logarithmic spin-dependence of the energy at fixed $g$ and increasing $S$.}
\label{fig:slow1}
}

\newpage
\FIGURE{
\epsfig{file=./der.eps, width=14cm}
\bigskip\bigskip
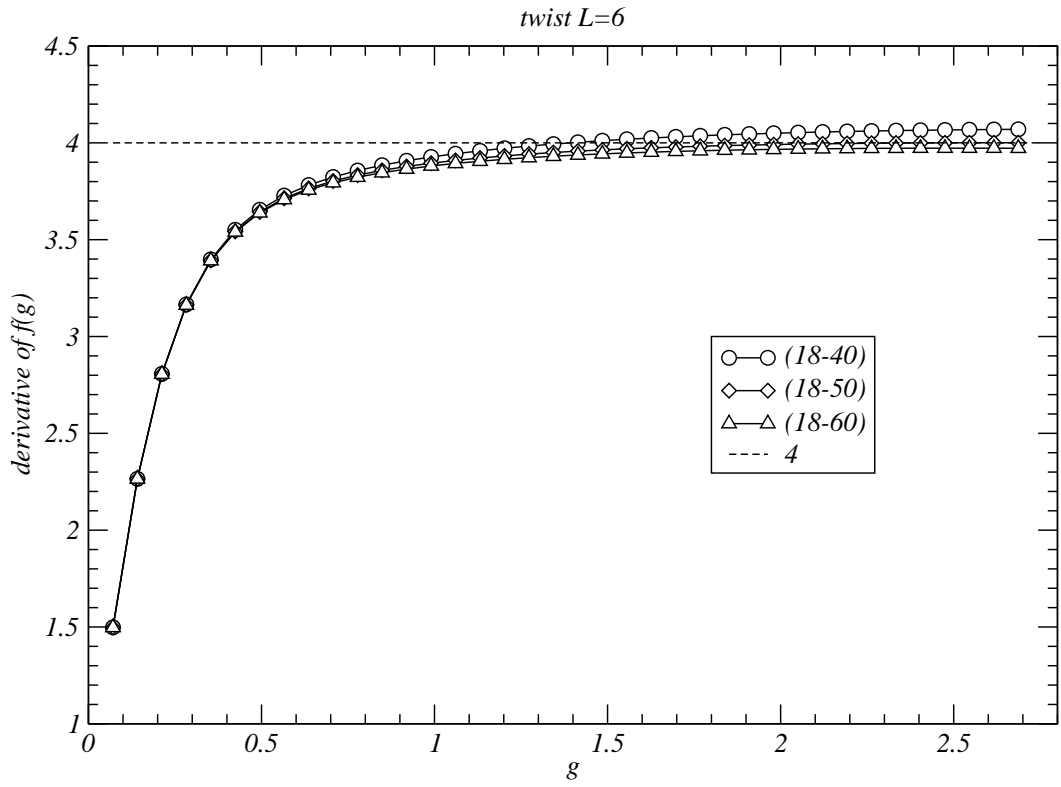
\caption{Long string limit. Best fit results for the asymptotic derivative of the scaling function.}
\label{fig:der}
}

%\newpage
%\FIGURE{
%\epsfig{file=Figures/spectrum.F4.eps,width=14cm}\bigskip\bigskip
%\caption{Smallest 6 energy levels at $F=4$ and various couplings $\lambda$ as functions of the upper limit on the boson
%number $B\le B_{\rm max}$.}
%\label{fig:spectrum}
%}

\end{document}